\documentclass[twocolumn,showpacs,preprintnumbers,amsmath,amssymb]{revtex4} 
\usepackage{graphicx} 

\usepackage{amsmath}
\usepackage{bm}
\usepackage{amssymb}

\begin{document}


\title{A spatially and temporally localized sub-laser-cycle electron source}

\author{Peter Hommelhoff, Catherine Kealhofer, and Mark A. Kasevich}

\affiliation{Physics Department, Stanford University, Stanford, CA
94305}

\date{June 30, 2006}

\begin{abstract}
We present an experimental and numerical study of electron
emission from a sharp tungsten tip triggered by sub-8 femtosecond
low power laser pulses. This process is non-linear in the laser
electric field, and the non-linearity can be tuned via the DC
voltage applied to the tip. Numerical simulations of this system
show that electron emission takes place within less than one
optical period of the exciting laser pulse, so that an 8 fsec 800
nm laser pulse is capable of producing a single electron pulse of
less than 1 fsec duration. Furthermore, we find that the
carrier-envelope phase dependence of the emission process is
smaller than 0.1\,\% for an 8 fsec pulse but is steeply increasing
with decreasing laser pulse duration.
\end{abstract}

\maketitle


Fast, laser driven electron sources are crucial as photo-electron
injectors for accelerators and free electron
lasers~\cite{Ayvazyan2002}. They have also enabled time-resolved
imaging of fundamental processes in such diverse fields as solid
state physics~\cite{Merano2005,Siwick2003},
chemistry~\cite{Ihee2001} and biology~\cite{Lobastov2005}. Due to
the nature of the emission processes utilized so far, the electron
pulses at best resemble the laser pulse {\em envelope} and are
usually more than 100~fsec long. Ultrafast sub-laser-cycle ($\sim
1$~fsec) electron sources have recently been obtained in
laser-ionization of gas phase atoms~\cite{Niikura2002}. However, a
spatially localized source with sub-laser-cycle resolution has
been unknown so far. Here we demonstrate a spatially and
temporally localized source of sub-1~fsec electron pulses. The
electron pulses are generated by the carrier electric field of a
three cycle laser pulse and are emitted from a sharp field
emission tip to confine the source area down to nanometric
dimensions. This electron source should find application in novel
optical accelerators~\cite{Plettner2005}, in new interferometers
and in all kinds of ultrafast electron microscopy~\cite{King2005}.


Our system consists of a field emission tip onto which we focus
sub-8 fs laser pulses with sufficiently large field strengths that
optical field emission dominates the emission process.  In
previous work, we have studied laser induced field emission using
longer ($\sim 70$~fs) laser pulses~\cite{Hommelhoff2006}. For such
long pulses, the electrons are expected to be emitted over a time
that is comparable to the length of the exciting laser pulse.
However, for few cycle laser pulses, optical field emission can
lead to electron emission in a time set by the laser {\em period}
so that the emitted electron pulse is significantly shorter than
the laser pulse itself. Optical field emission from solids is the
direct analog to optical tunnel ionization in
atoms~\cite{Brabec2000}, with the distinct advantage that the
emitted electrons originate from an area determined by the tip
dimension (on the order of nanometers) and not by the laser spot
size (typically microns), and that the electron beam from a field
emission tip is well collimated down to about $10^{\circ}.$

Considered from a different point of view, the electron emission
processes described above can be used to probe the exciting laser
{\em electric field}, in stark contrast to more conventional
sensors that probe the laser intensity and average over at least
one optical period in the laser field. This distinction becomes
important for laser pulses shorter than three optical periods, and
critical for pulses that are close to or in extreme cases even
shorter than one optical period~\cite{Shverdin2005}. The value of
the electric field is usually parameterized by the
carrier-envelope phase angle, which describes the phase advance of
the electric carrier field with respect to the pulse
envelope~\cite{Brabec2000}. In this work, we study optical field
emission from a tungsten tip both from the perspective of
generating ultrashort electron pulses and from the perspective of
developing a simple carrier-envelope phase detector.


\begin{figure}[htb]
\centerline{\includegraphics[width=2.8cm]{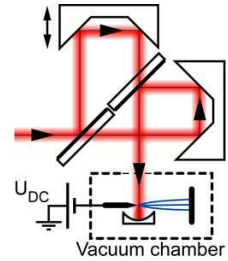}}
\caption{\label{figSetup} (Color online). Sketch of the
experimental setup. The laser beam traverses a dispersion balanced
interferometer, whose output is focused onto the field emission
tip.}
\end{figure}

In our setup a tungsten tip in (111) orientation with a radius of
curvature of around $80\,$nm is mounted in an ultrahigh vacuum
chamber (Fig.\,\ref{figSetup}). A Kerr-lens mode-locked
Ti:Sapphire laser generates laser pulses with a repetition rate of
150\,MHz and an average output power of around 500\,mW. We measure
the pulse duration with a standard doubling crystal autocorrelator
to be below but close to 8~fs. The laser beam traverses a
dispersion balanced interferometer (Fig.\,\ref{figSetup}), which
allows the measurement of interferometric autocorrelation traces
using the field emission tip as a non-linear element.  The output
beam from the interferometer is focused on the tip with a
spherical gold mirror to a spot radius of $\sim 4\,\mu$m
($1/e^2$).  The tip is oriented perpendicular to the laser beam
direction, and the polarization vector of the laser light is
parallel to the tip shank. Emitted electrons are accelerated
towards an imaging microchannel plate detector and the amplified
current is measured. In addition to the laser electric field, it
is possible to add a DC field to the tip by applying a voltage
$U_{\mathrm{DC}}.$ For sharp tips, the electric field $F$ and the
applied voltage $U$ are related by the equation $F = U / (kr),$
with the tip radius $r$ and the geometric factor $k \approx 5$
\cite{Binh1996}.

With an average laser power of $\sim 100\,$mW, electric field
strengths of around 0.75\,GV$/$m can be generated in the focal
spot. Because the dimensions of the tip are smaller than the laser
wavelength, AC field enhancement takes place. We expect an
enhancement factor of around 4 so that the electric field at the
tip can reach 3\,GV$/$m \cite{Martin2001, Hommelhoff2006}. In
previous work, we have shown that at these field strengths optical
field emission dominates electron emission~\cite{Hommelhoff2006}.

\begin{figure}[htb]
\centerline{\includegraphics[width=8.6cm]{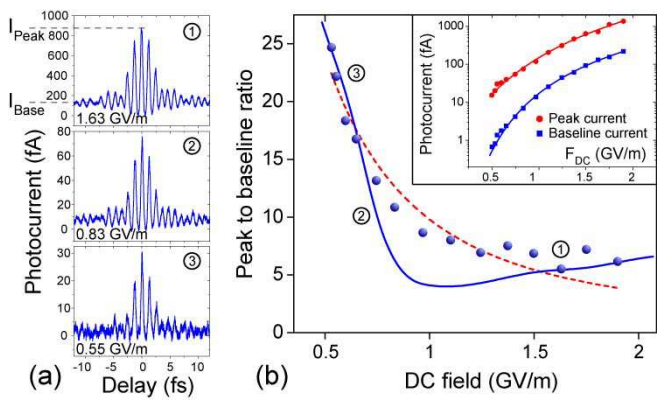}}
\caption{\label{figPeakBaselineRatio} (Color online).
Autocorrelation traces with tunable non-linearity. (a)
autocorrelation traces for three different DC voltages but
identical laser parameters. (b) peak-to-baseline ratio versus DC
tip voltage (blue points: data). The solid blue line is the
simulation result with no adjustable parameter, the dashed red
line is a fit of the simple optical field emission model to the
data. The numbers in circles correspond to the traces given in
(a). Inset: Peak current and baseline current versus DC tip
voltage.}
\end{figure}

Fig.\,\ref{figPeakBaselineRatio}(a) shows three interferometric
autocorrelation traces (IATs) recorded in the electron current for
three different DC electric fields but identical laser parameters.
We observe that the emission current for two pulses that are
separated by $\sim 100~$fs is additive. This rules out the
possibility that the non-linearity in the emission process is due
to thermionic emission. Thermionic emission scales exponentially
with the deposited laser energy~\cite{Binh1996} and has typical
time scales of 100~fs to 1~ps~\cite{Riffe1993}. Adjusting the DC
field causes the shape of the IAT's to change because the
non-linearity of the emission process increases in a continuous
manner as the DC field is reduced. To quantify this, we divide the
peak emission current by the baseline current and plot this ratio
versus the DC field [Fig.\,\ref{figPeakBaselineRatio}(b)]. For
$F_{\mathrm{DC}} = 0.53\,$GV$/$m the peak-to-baseline ratio
reaches 25, which should be compared with peak-to-baseline ratios
of 8 and 32 for conventional second and third order processes.

The dashed curve shows a semi-quantitative model for the data
based on optical field emission. In the optical field emission
regime, the tunnel current at a given time can be viewed as that
arising from DC tunnelling with the electric field given by the
sum of the applied DC field and the instantaneous value of the
laser field. The DC tunnelling is described by the Fowler-Nordheim
equation~\cite{Binh1996}, which relates the tunnel current $I$ to
the applied electric field $F$:
\begin{equation} \label{eqFN} I = A F^2 \exp\left(-\frac{B}{F}\right)
\end{equation} with $A$ and $B$ about constant.  Due to the exponential
non-linearity in this equation, the photo-current obtained at the
maximum of the laser electric field ($F_{\mathrm{laser}}$) makes
the largest contribution to the total emission probability. This
is given by
\begin{equation} \label{eqTwoPulsesNonOverlapped}
I_{\mathrm{Base}} = 2A (F_{\mathrm{laser}} + F_{\mathrm{DC}})^2
\exp\left(-\frac{B}{F_{\mathrm{laser}} + F_{\mathrm{DC}}}\right),
\end{equation}
 for two non-overlapping laser pulses and
 \begin{equation}
\label{eqFNTwoPulsesOverlapped} I_{\mathrm{Peak}} = A
\left(2F_{\mathrm{laser}} + F_{\mathrm{DC}}\right)^2
\exp\left(-\frac{B}{2F_{\mathrm{laser}} + F_{\mathrm{DC}}}
\right).
\end{equation}
for two perfectly overlapping identical pulses. We fit the ratio
of equation~(\ref{eqFNTwoPulsesOverlapped}) and
equation~(\ref{eqTwoPulsesNonOverlapped}) to the data of
Fig.\,\ref{figPeakBaselineRatio} with $F_{\mathrm{laser}}$ as free
parameter. The model fits the data reasonably well and we obtain
as best fit value $F_{\mathrm{laser}} = (1.8 \pm 0.2)\,$GV$/$m,
which is consistent with the parameters of our system given the
uncertainty in the size of the laser focus and field enhancement
factor.

\begin{figure}[htb]
\centerline{\includegraphics[width=8cm]{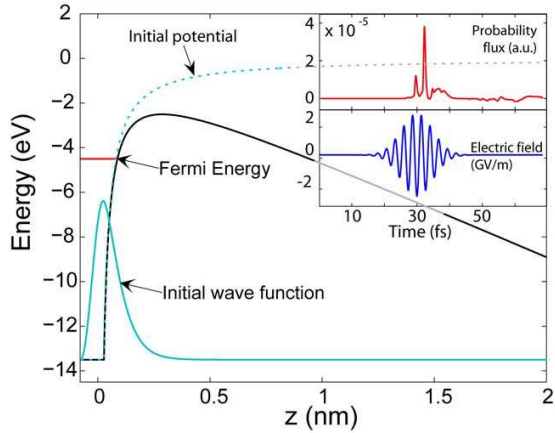}}
\caption{\label{figInitialState}(Color online). Model and electron
pulse. Model potential used in the calculations, shown here with
an applied electric field of $F= 4.4$~GV/m (solid black line).
$E=0$ is the vacuum energy. Initial potential and  initial
wavefunction are shown in light blue. The inset shows the electric
field generated by a three cycle laser pulse ($F_{\mathrm{laser}}
\approx 2.7$\,GV/m, $F_{\mathrm{DC}} = 0.2$\,GV/m, CE phase $\phi
= \pi,$ bottom) and the resulting probability flux (top). A single
electron pulse with a width of 660\,as (FWHM) is emitted. The
measurement position was at $z=2$\,nm for this plot.}
\end{figure}


The Fowler-Nordheim theory used above describes field emission in
steady state for a DC field. For time dependent processes, in
particular for processes involving laser pulses with durations
comparable to electronic time scales in metals, this model might
prove to be insufficient. In order to include key dynamical
effects, we have integrated the one-dimensional time-dependent
Schr\"odinger equation. The potential is modelled as shown in
Fig.~\ref{figInitialState}, where the size of the box is chosen
such that the ground state wavefunction matches the electronic
Fermi energy in tungsten. This model reflects the fact that field
emission from (111)~planes should be dominated by emission from
surface states~\cite{Ohwaki2003}. The effect of the laser field is
to modulate the barrier so that the potential reads $V(z, t) = V_0
= -13.5~$eV inside the metal $(z \le 0)$ and $V(z, t) = -q^2 / (16
\pi \epsilon_0 z) - q z [F_{\mathrm{DC}} + F_0 \exp (-2 \ln 2 t^2/
\tau^2) \cos (\omega t + \phi)]$ outside. Here, the first term is
the image potential, $F_0$ is the peak of the laser electric field
envelope, $F_{\mathrm{DC}}$ is the applied DC field, $z$ the
spatial coordinate and $\phi$ the carrier-envelope phase angle. We
find the initial ground state wave function (in absence of the
tunnel barrier: $F_0 = F_{\mathrm{DC}} = 0$) with the imaginary
time method. The simulation records the probability flux at $z
\approx 6$~nm outside of the metal for each time step. With this
method, we calculate the probability current integrated over a
single laser pulse as a function of pulse duration, pulse energy,
DC tip voltage and carrier-envelope phase.

\begin{figure}[htb]
\centerline{\includegraphics[width=7cm]{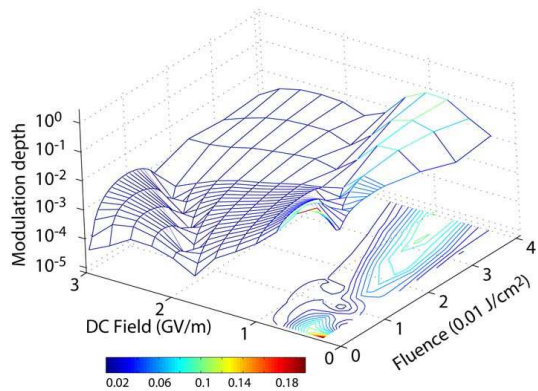}}
\caption{\label{figModDepth3D}(Color online). Modulation depth as
a function of DC field and fluence for a 2 cycle laser pulse
($\tau = 5.3$\,fs). The surface plot is on a log scale, whereas
the contour plot is on a linear scale (colour bar applies to the
latter). The two distinct valley structures might reflect
dynamical resonance effects.}
\end{figure}

From tunnel current vs. fluence behavior we extract the
non-linearity and calculate the expected peak to baseline ratio.
The result is the solid blue line in
Fig.~\ref{figPeakBaselineRatio}, a theory curve with no adjustable
parameters.  Given that our model does not include any properties
of the emitter material other than the work function, the
agreement with the data is very good. From this and also from the
agreement of the simple model described earlier we infer that it
is optical field emission that dominates the emission process. The
slight deviations in both models might be due to competing
processes, owing to the fact that the Keldysh parameter in our
experiment is of order 1.

In principle, optical field emission is extremely sensitive to the
electric field of the laser pulse.  Because of the exponential in
equation~(\ref{eqFN}), optical tunnel emission was the prominent
process envisioned to enable a direct measurement of
carrier-envelope effects~\cite{Xu1996}. Although nowadays several
sensors for the CE phase of ultrashort laser pulses
exist~\cite{Paulus2003,Apolonski04,Fortier04}, none of these
advances the quest to obtain direct CE phase locking of the laser.
Miniaturized, a detector like the one used in this work could be
used as such a system, with the observable being a modulation in
the emission current depending on $\phi.$ Fig.~\ref{figModDepth3D}
shows the modulation depth [minimum current (as function of
$\phi$) subtracted from maximum current divided by the sum] as a
function of pulse energy and DC field for a two-cycle pulse. The
largest modulation depth (21\,\%) lies close to the origin and
stays on that order of magnitude on a ridge that runs under a
small angle to the fluence axis.

\begin{figure}[htb]
\centerline{\includegraphics[width=8cm]{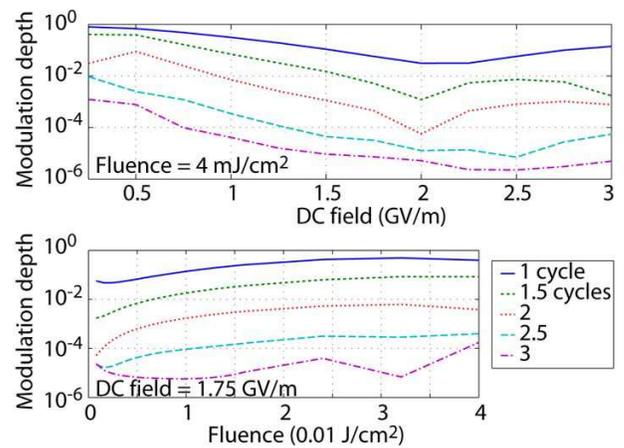}}
\caption{\label{figModDepth2D}(Color online). Modulation depth as
a function of DC field and fluence for different pulse durations.}
\end{figure}

Fig.~\ref{figModDepth2D} displays the modulation depths for laser
pulses with different pulse durations. 3D plots for each pulse
duration show similar behaviors to the one shown in
Fig.~\ref{figModDepth3D}. We find that for a three cycle pulse the
carrier-envelope phase modulates the emission current by no more
than $0.1~\%.$ In contrast, a time-independent Fowler-Nordheim
calculation yields an almost one order of magnitude larger
modulation depth. We have studied the problem experimentally by
locking the carrier envelope frequency of our laser using an f-2f
interferometer~\cite{Holzwarth2000} and looking for a signal at
the carrier envelope frequency in the emitted electron current.
With this method, we find that the modulation depth is indeed
smaller than $\sim 0.1~\%$ for a three cycle laser pulse and so
confirm the results of the time-dependent calculation. The
discrepancy between the time-dependent and the time-independent
calculation presumably stems from the fact that the barrier
modulation timescale is not much longer than the electronic
timescales within the metal. We conclude that a field emitter
based sensor is less suitable than initially
expected~\cite{Xu1996}. However, based on the simulation results,
we believe that for pulses shorter than $\sim 2$ optical periods a
field emitter based system will represent a good carrier envelope
phase detector. Future work will seek to provide experimental
confirmation.

In optical field emission, individual laser cycles are expected to
be resolved. In fact, in the simulation we observe sub-laser-cycle
electron pulses for a wide range of parameters (inset in
Fig.~\ref{figInitialState}). For our experimental conditions, this
corresponds to electron emission times of around 660\,as. Because
CE phase stabilized lasers have been shown to possess a timing
jitter of down to 40\,as \cite{Fuji2005, Mucke2005}, the timing
jitter of the electron emission lies well in the attosecond
domain. Evidently, a finite initial energy spread and Coulomb
repulsion will lead to a loss of timing accuracy. Fast
acceleration of the electrons to energies $\gtrsim 100\,$keV, a
limitation to one electron per pulse~\cite{Lobastov2005}, and
laser-induced dispersion control will mitigate these effects. An
electron source with such a high time-resolution would be
extremely desirable for future laser accelerators, in which the
commonly used microwave acceleration field will be replaced by an
optical frequency electric field~\cite{Plettner2005}, and photonic
crystal waveguides will take the place of microwave
cavities~\cite{Lin2001,Cowan2003}. Additionally, time-resolved
imaging of biological, chemical and solid state processes with
faster and brighter electron sources will push forward knowledge
in each respective field~\cite{Ihee2001, Siwick2003, Merano2005}.

We would like to thank Steve Harris's group and MenloSystems for
lending us equipment and Phil Bucksbaum for discussions. This work
was supported by grants from the ARO MURI program and by the
Humboldt Foundation (P.H.).


\end{document}